\documentclass[prl, twocolumn, superscriptaddress, showpacs, nobibnotes, reprint,floatfix]{revtex4-1}

\usepackage{graphicx}
\usepackage{epstopdf}
\usepackage{amsmath,textcomp}

\def\ket #1{\vert #1\rangle}

\def\us {$\mu$s}
\def\uphi {$\mu\Phi_0$}
\def\rthz {$/\sqrt{\mathrm{Hz}}$}
\def\nbar {$\bar{n}$}

\def\nbarsp {$\bar{n}$ }
\def\ff {$\omega_\mathrm{ff}$}
\def\pff {$P_\mathrm{ff}$}

\newcommand{\beq}{\begin{equation}}
\newcommand{\eeq}{\end{equation}}

\begin{document} 

\title{Measurement-induced qubit state mixing in circuit QED from up-converted dephasing noise}
\author{D. H. Slichter}
\email[Electronic address: ]{slichter@berkeley.edu}
\altaffiliation[Present address: ]{Time and Frequency Division, National Institute of Standards and Technology, Boulder CO 80305}
\author{R. Vijay}
\author{S. J. Weber}
\affiliation{Quantum Nanoelectronics Laboratory, Department of Physics, University of California, Berkeley CA 94720}
\author{S. Boutin}
\author{M. Boissonneault}
\affiliation{D\'epartement de Physique, Universit\'e de Sherbrooke, Sherbrooke, Qu\'ebec, Canada J1K 2R1}
\author{J. M. Gambetta}
\affiliation{IBM T.J. Watson Research Center, Yorktown Heights, NY 10598}
\author{A. Blais}
\affiliation{D\'epartement de Physique, Universit\'e de Sherbrooke, Sherbrooke, Qu\'ebec, Canada J1K 2R1}
\author{I. Siddiqi}
\affiliation{Quantum Nanoelectronics Laboratory, Department of Physics, University of California, Berkeley CA 94720}

\date{\today}

\begin{abstract}
We observe measurement-induced qubit state mixing in a transmon qubit dispersively coupled to a planar readout cavity.  Our results indicate that dephasing noise at the qubit-readout detuning frequency is up-converted by readout photons to cause spurious qubit state transitions, thus limiting the nondemolition character of the readout. Furthermore,  we use the qubit transition rate as a tool to extract an equivalent flux noise spectral density at $f \sim 1$ GHz and find agreement with values extrapolated from a $1/f^\alpha$ fit to the measured flux noise spectral density below 1 Hz.  
\end{abstract}


\maketitle

High-fidelity measurement is a crucial tool in quantum information science.  For superconducting qubits \cite{Siddiqi2011, Clarke2008}, one widely used framework for performing quantum nondemolition (QND) \cite{Braginsky1992} measurement is the circuit quantum electrodynamics (cQED) architecture \cite{Blais2004, Wallraff2004}.  In cQED, a qubit is coupled to a microwave-frequency resonant cavity through a Jaynes-Cummings-type interaction, in analogy to an atom in an optical Fabry-Perot cavity.  In the dispersive limit, probing the qubit-state-dependent resonant frequency of the cavity implements, to first order, a QND measurement of the qubit state.  

In the case of a linear readout cavity \cite{Wallraff2005}, only recently has single-shot sensitivity been demonstrated using a near-quantum-noise-limited superconducting parametric amplifier \cite{Hatridge2011, Levenson-Falk2011}, enabling observation of individual qubit state transitions in real time \cite{Vijay2011}.  Subsequent experiments \cite{Johnson2012, Riste2012} have reported single-shot fidelities of 94\%-97\%. Nonlinear circuit QED readout methods\textemdash using either the nonlinearity of the qubit \cite{Reed2010a, Boissonneault2010, Bishop2010} or a nonlinear cavity \cite{Mallet2009}\textemdash have shown single-shot fidelities of 86\%-92\%, but the former is not QND and the latter is too slow to allow continuous qubit monitoring.

In this letter, we explore non-QND behavior in cQED readout with a linear cavity. We employ single shot readout \cite{Vijay2011} to directly quantify the rate of measurement-induced qubit transitions. We find that dephasing noise at the qubit-readout detuning frequency $\Delta_\mathrm{ro}=\omega_q-\omega_\mathrm{ro}$ combines with readout photons to induce qubit excitation and relaxation, making the measurement process no longer fully QND.  The rate of qubit transitions due to such ``dressed dephasing'' depends linearly on the average cavity photon occupation \nbarsp and the spectral density of dephasing noise at the detuning frequency $S(\pm\Delta_\mathrm{ro})$, consistent with recent calculations which keep higher order terms in the dispersive approximation \cite{Boissonneault2009}. Furthermore, the qubit transition rate provides a new probe of dephasing noise at $|\Delta_\mathrm{ro}|/2\pi\sim 1$ GHz, a frequency range not currently accessible by other techniques.  We find that our extracted value of dephasing noise at GHz frequencies is consistent with the ``universal" 1/f magnetic flux noise \cite{Wellstood1987, Koch2007a} typically observed in low frequency measurements, suggesting the persistence of this noise mechanism over 11 orders of magnitude in frequency.

Dephasing can be described with the Hamiltonian
\beq
\label{dephham}
H_{\varphi}=\hbar\nu f_{\varphi}(t)\hat{\sigma}_{z},
\eeq
where $f_{\varphi}(t)$ is a random noise (e.g. flux noise) with zero mean and $\nu$ characterizes the coupling between the noise and the qubit.  We take $\nu f_{\varphi}(t)$ to be a small perturbation on the qubit frequency $\omega_q$.  While the low-frequency ($\ll \omega_q$) components of $f_{\varphi}(t)$ are typically the dominant source of qubit dephasing in experiments, the frequency spectrum of $f_{\varphi}(t)$ can also have components at $\Delta_\mathrm{ro}$, which can combine with readout photons to cause transitions between the qubit states.  The rate for transitions up and down due to this dressed dephasing is given by \cite{Boissonneault2009,Boissonneault2012}
\beq
\label{ddeq1}
\Gamma_{\uparrow\!\downarrow,\mathrm{DD}}=4\frac{g^{2}}{\Delta_\mathrm{ro}^{2}}\nu^{2}S(\mp\Delta_\mathrm{ro})\bar{n},
\eeq
where $g$ is the qubit-cavity coupling, $\bar{n}$ is the average cavity photon occupation, and $S(\Delta_\mathrm{ro})$ is the power spectral density of $f_{\varphi}(t)$ at the detuning frequency $\Delta_\mathrm{ro}$. This expression holds for $\bar{n}\ll n_\mathrm{crit}=\Delta_\mathrm{ro}^2/4g^2$.  In the case of a symmetric noise spectrum where $S(\Delta_\mathrm{ro})=S(-\Delta_\mathrm{ro})$, we have $\Gamma_{\downarrow,\mathrm{DD}}=\Gamma_{\uparrow,\mathrm{DD}}$.  Once the system has reached steady state, we expect a spurious excited state population to exist.  Using the principle of detailed balance, we can express this as 
\beq
\label{ddeq2}
\langle\hat{\sigma}_{z}\rangle=-1+\frac{\Gamma_{\uparrow,\mathrm{DD}}+\Gamma_{\uparrow,\mathrm{th}}}{\Gamma_{1}+\Gamma_{\downarrow,\mathrm{DD}}+\Gamma_{\downarrow,\mathrm{th}}}\approx-1+\frac{\Gamma_{\uparrow,\mathrm{DD}}}{\Gamma_{1}} +\frac{\Gamma_{\uparrow,\mathrm{th}}}{\Gamma_{1}}\, .
\eeq
Here $\Gamma_1=1/T_1$ is the intrinsic qubit decay rate, including the Purcell effect, and $\Gamma_{\uparrow\!\downarrow,\mathrm{th}}$ is the qubit's thermal excitation/relaxation rate [the rightmost term of Eq. (\ref{ddeq2}) gives the average thermal population of the qubit].  We have used the approximation that $\Gamma_{\uparrow\!\downarrow,\mathrm{th}}, \Gamma_{\uparrow\!\downarrow,\mathrm{DD}}\ll \Gamma_1$, which is valid for our experimental conditions.

Our experiment, shown schematically in Figure 1, is anchored to the mixing chamber of a dilution refrigerator at 50 mK.  A transmon qubit \cite{Koch2007}  ($E_{J,\mathrm{max}}$=21.7 GHz, $E_C$=220 MHz) is capacitively coupled ($g/2\pi=106$ MHz) to a planar superconducting quasi-lumped-element readout cavity \cite{Khalil2011, Geerlings2012} consisting of a meander inductor ($L=3.7$ nH) in parallel with an interdigital capacitor ($C=175$ fF), giving a bare resonant frequency of 6.2724 GHz.  The cavity has asymmetric coupling and is operated in transmission; the strongly coupled port sets the cavity linewidth $\kappa/2\pi=7$ MHz.  Qubit manipulation and readout signals enter from the weakly coupled port via a heavily attenuated injection line.  Readout photons in the cavity acquire a phase shift that depends on the state of the qubit, then leave through the strongly coupled port and are amplified by a superconducting parametric amplifier (paramp) \cite{Vijay2011,Slichter2011}. Four microwave circulators isolate the qubit from the strong paramp pump tone.  Further amplification is performed by cryogenic and room temperature amplifiers (not shown). The output signal is finally detected by homodyne mixing and then digitized.  This method allows us to monitor the qubit state in real time and record quantum jumps between qubit energy levels, as seen in the inset figure.

In addition, a weakly coupled fast flux line allows modulation of the qubit Hamiltonian by noise or coherent signals.  The fast flux line has a bandwidth of 2.2 GHz, defined by a reactive filter at 100 mK and three lossy impedance-matched low-pass filters \cite{Slichter2009} at 4 K, 100 mK, and 50 mK.  These filters thermalize the line without introducing excessive low frequency loss, allowing us to pass large currents in the fast flux line without heating the mixing chamber.  To calibrate the coupling of the fast flux line to the qubit loop, we extract the flux-to-qubit-frequency transfer function from qubit spectroscopy.   We then measure the qubit frequency as a function of applied dc current through the fast flux line using Ramsey fringes.  Combining these factors with the measured frequency-dependent attenuation of the fast flux line allows us to convert room-temperature power into a flux in the qubit loop.  The coupling is sufficiently weak (120 mA/$\Phi_0$) that Johnson noise from the 50 $\Omega$ impedance of the fast flux line is not the dominant source of flux noise in the qubit loop for frequencies at or below $\omega_q/2\pi$.  

\begin{figure}[tbp]
\begin{center}
\includegraphics{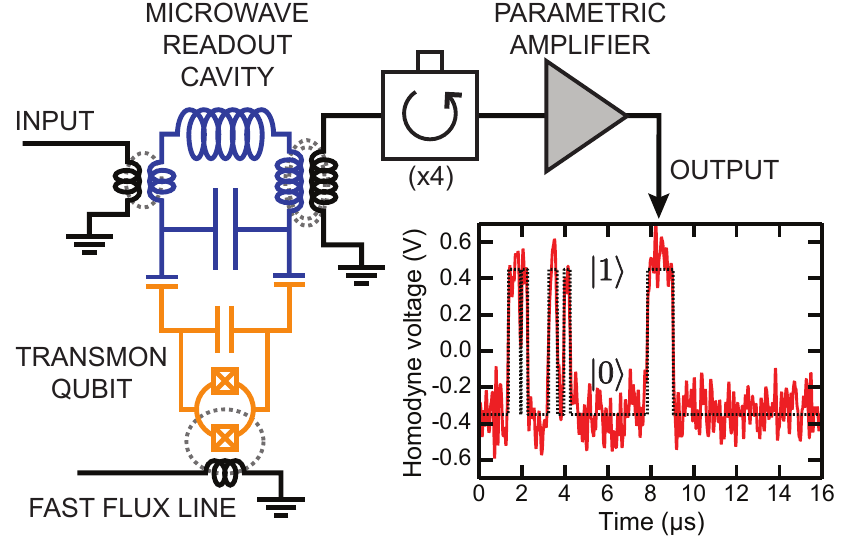}
\end{center}
\caption{Experimental setup and readout trace (color online).  The transmon qubit is coupled to an asymmetric microwave readout cavity.  Noise or coherent tones can be injected into the qubit loop via a weakly coupled fast flux line with 2.2 GHz bandwidth (grey dashed lines indicate flux coupling).  The readout signal is amplified by a superconducting paramp, enabling continuous high-fidelity monitoring of the qubit state.  The inset shows a sample output trace (red line) and the corresponding extracted qubit state (dotted black line).}
\end{figure}

For each qubit bias point, the dispersive shift $2\chi$ is calculated using an expression that accounts for the higher excited states of the transmon qubit and includes corrections for the qubit-induced Kerr nonlinearity \cite{Boissonneault2010}.  We use the dispersive shift information to calibrate the number of photons in the readout cavity using the ac Stark shift \cite{Schuster2005} and to choose the readout frequency $\omega_\mathrm{ro}/2\pi$.  We select a readout frequency halfway between the cavity resonant frequencies corresponding to the qubit in the ground and first excited states $\omega_\mathrm{ro}=\frac{1}{2}[\omega_\mathrm{cav}(\ket{0})+\omega_\mathrm{cav}(\ket{1})]=\omega_\mathrm{cav}(\ket{0})+\chi$.  This choice of readout frequency simplifies our analysis because the average cavity photon occupation \nbarsp is unaffected (up to a 5-10\% correction at the highest \nbarsp used in the experiment) by whether the qubit is in state $\ket{0}$ or $\ket{1}$.  Qubit coherence times varied monotonically depending on the qubit frequency, from $T_1=290$ ns and $T_2^*=550$ ns  at $\omega_q/2\pi=5.705$ GHz to $T_1=910$ ns and $T_2^*=1.35$ \us\, at $\omega_q/2\pi=5.075$ GHz.  These numbers represent $T_1$ values about a factor of two below the Purcell limit, and pure dephasing times $T_\varphi$ much longer than $T_1$.  

The measurement protocol consisted of readout pulses lasting 17.5 \us \;occurring every 100 \us.  The long delay ensured that the qubit would fully relax to its thermal ground state ($\sim$1.4\% excited state population, corresponding to a qubit temperature of 60 mK) between measurement runs.  We took $10^4$ individual time traces for each combination of experimental parameters.  The measurement traces were analyzed by smoothing to optimize signal-to-noise ratio and then using a hysteretic threshholding algorithm similar to that demonstrated in \cite{Yuzhelevski2000} to determine the qubit state at each time point.  This allowed qubit state populations to be determined both at steady state and as a function of time into the readout.  The population extraction algorithm was tested on simulated data traces with realistic experimental parameters, and shown to give fractional errors of less than 5\% for population estimates \cite{Slichter2011}.

\begin{figure}[tbp]
\includegraphics{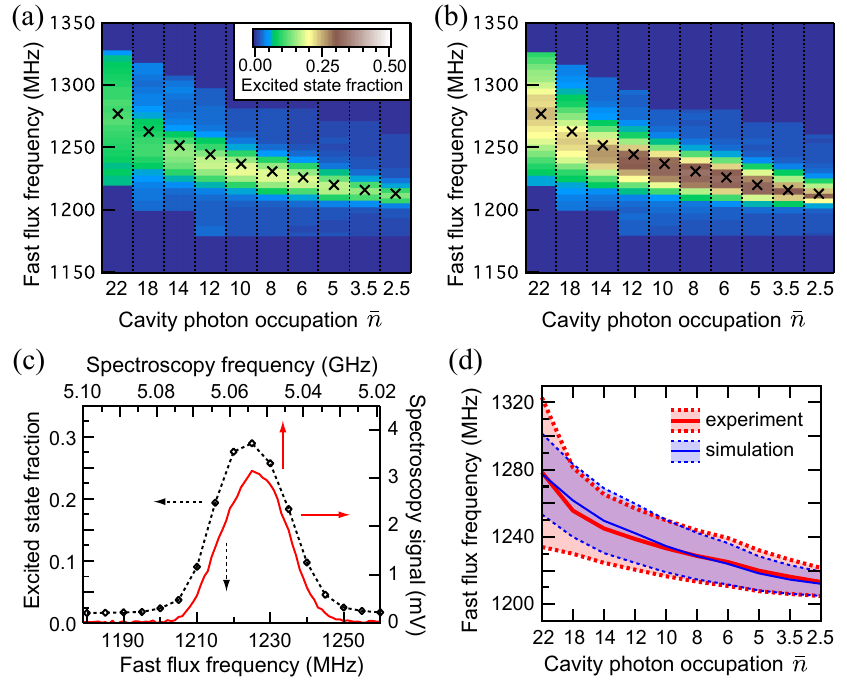}
\caption{\label{tonefig}Spurious excitation with coherent fast flux tone (color online).  The qubit steady state population during measurement with a coherent microwave tone applied to the fast flux line is shown as as a function of \ff\! and \nbar.  The panels correspond to RMS added fluxes of 300 \uphi\! (a) and 550 \uphi\! (b).  The black x's denote the values of $\Delta_\mathrm{ro}(\bar{n})$ extracted independently from qubit spectroscopy.  Regions with no data (at the lowest and highest frequencies) are shown in uniform dark blue.  Panel (c) shows the qubit population versus \ff\! for $\bar{n}=6$ and an RMS added flux of 420 \uphi\! (black dotted line and diamonds), along with the corresponding qubit spectroscopy trace (red solid line).  The horizontal axes are aligned such that the detuning between readout and spectroscopy frequencies is the same as the fast flux frequency.  Panel (d) shows the center (solid line) and widths (dotted lines) of the qubit excited state population response for the data in (b).  The red (bold) traces are experimental data, while the blue (narrow) traces are from numerical simulations.}
\end{figure}

To test the dressed dephasing theory, we began by injecting a continuous microwave tone into the fast flux line at a frequency \ff\,$\sim|\Delta_\mathrm{ro}(\bar{n})|$, where we note that $\Delta_\mathrm{ro}$ depends on \nbar\, due to the ac Stark shift \cite{Schuster2005}.  This tone produces a small flux oscillation in the qubit loop; the power \pff\! of the tone was varied to produce RMS flux excitations of up to 825 \uphi, corresponding to qubit frequency fluctuations of up to 3-6 MHz RMS (depending on qubit bias point).  These fluctuations are much smaller than the ac-Stark-broadened qubit linewidth for the values of \nbar \, studied.  For each value of \nbar, we stepped the \ff\! through a range of around 100 MHz centered on $|\Delta_\mathrm{ro}(\bar{n})|$.  

The results of this experiment (with the qubit biased at $\omega_q/2\pi= 5.075$ GHz) are shown in Figure \ref{tonefig}.  When \pff\,$\neq 0$, qubit state mixing occurs as long as \ff\! is within roughly a qubit linewidth of the detuning frequency, and is most noticeable when \ff\,=$|\Delta_\mathrm{ro}(\bar{n})|$.  Panels (a) and (b) show the qubit excited state population with varying \nbarsp for \pff=300 \uphi\! and 550 \uphi, respectively.  The value of $|\Delta_\mathrm{ro}(\bar{n})|$, found independently from qubit spectroscopy, is denoted with black crosses.  Figure \ref{tonefig}(c) plots the spurious excitation as a function of \ff\! along with the independently measured spectroscopy signal for the same \nbar; the horizontal axes are aligned such that the fast flux frequency is the same as the detuning between the readout and spectroscopy frequencies.  This choice highlights the correlation between the values of \ff\! which cause qubit state mixing and $\Delta_\mathrm{ro}$.  

Figure \ref{tonefig}(d) shows the extracted center frequency and linewidth of the regions of qubit excitation in Fig. \ref{tonefig}(b).  We also plot the center frequency and linewidth of qubit excitation, for the same experimental parameters, calculated from numerical simulations of the multilevel Jaynes-Cummings Hamiltonian with an added coherent flux tone.  The simulations agree well with the experimental data on the location and width of the peak, with some width discrepancy appearing at high \nbar.  The simulated steady-state excited populations agree with experiment to within a factor of $\sim$1.4 or better, depending on the qubit bias parameters.  We attribute the remaining discrepancy to uncertainty in the calibration of our fast flux line.  

In the absence of a fast flux tone, we still observe some spurious qubit excitation with increasing \nbar, which we postulate is due to the intrinsic qubit flux noise at $\Delta_\mathrm{ro}(\bar{n})$ being up-converted by readout photons.  The effect is about 1\% additional excited population at steady state per 10 photons cavity occupation, increasing more rapidly at higher photon numbers (up to 10-15\% excited state population for $\bar{n}\approx 40$).  This measurement-induced state mixing, which reduces the fidelity of cQED measurement, can also be used as a spectrometer for dephasing noise at $\Delta_\mathrm{ro}$.  

\begin{figure}[tbp]
\begin{center}
\includegraphics{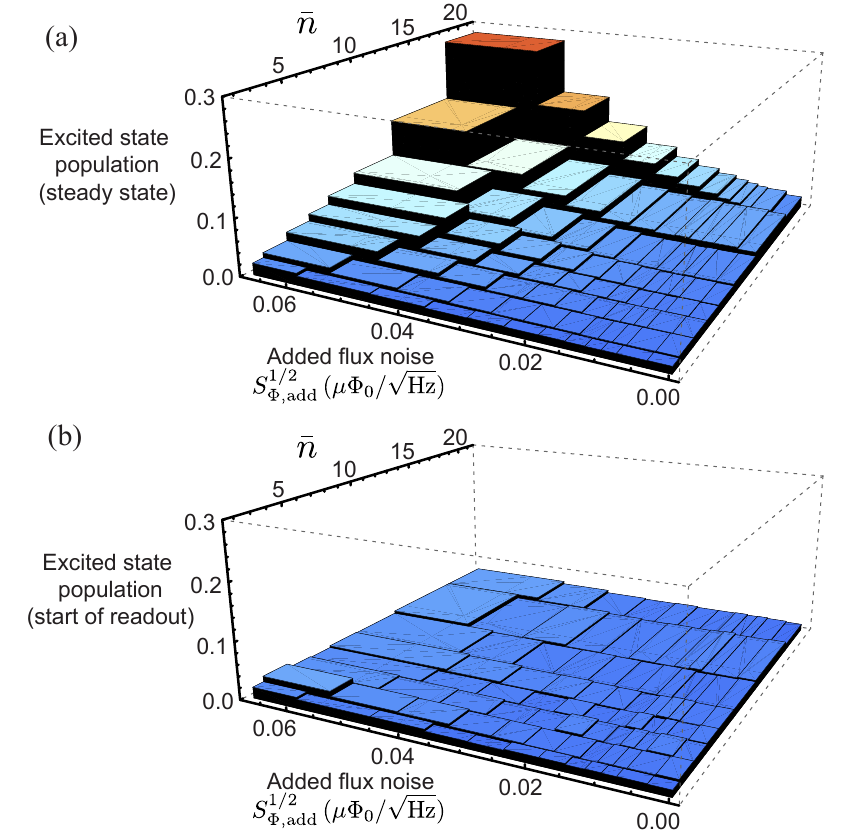}
\end{center}
\caption{\label{noisefig}Qubit population with added noise (color online). Part (a) shows the steady-state qubit population as a function of added flux noise at $|\Delta_\mathrm{ro}(\bar{n})|$ and the number of measurement photons \nbar \, in the readout cavity.  Part (b) shows the qubit population immediately after the readout has fully energized.  A thermal population of 1.4\% is visible, and is equivalent for all values of added flux noise and \nbar.}
\end{figure}

We examined this notion by intentionally applying flux noise to the qubit loop using the fast flux line and observing spurious excitation during measurement.  The noise was generated by amplifying the Johnson noise of a room-temperature 50 $\Omega$ termination.   The experiment was performed with white noise filtered to lie either in the band from 10 MHz to 2.2 GHz or from 180 MHz to 2.2 GHz.  The steady-state qubit populations were essentially identical between these two types of applied flux noise, suggesting again that only dephasing noise components near $\Delta_\mathrm{ro}(\bar{n})$ are responsible for spurious excitation.  

Figure \ref{noisefig}(a) shows the steady-state qubit excited population as a function of \nbarsp and the spectral density of added flux noise at the detuning frequency $S_{\Phi,\mathrm{add}}^{1/2}[\Delta_\mathrm{ro}(\bar{n})]$.  The qubit excited state population scales roughly linearly in \nbar\, and quadratically in $S_{\Phi,\mathrm{add}}^{1/2}[\Delta_\mathrm{ro}(\bar{n})]$, as predicted by the dressed dephasing theory.  To ensure that the added noise is not causing excitation in the absence of measurement, we also examine the qubit population at the start of measurement from the same data set, shown in Figure \ref{noisefig}(b).  We find that the excited state population at the start of the measurement corresponds to the thermal population, and is independent of \nbar\, and $S_{\Phi,\mathrm{add}}^{1/2}[\Delta_\mathrm{ro}(\bar{n})]$.

\begin{figure}[tbp]
\begin{center}
\includegraphics{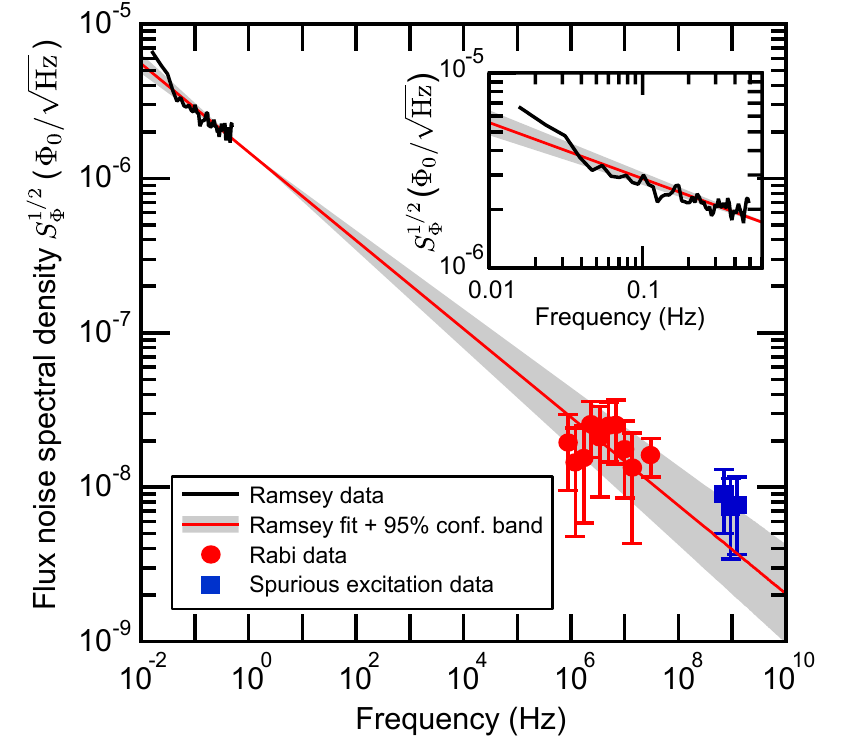}
\end{center}
\caption{\label{slopefig}Spectral density of flux noise vs frequency (color online).  We plot equivalent flux noise extracted from Ramsey fringes (black line), Rabi oscillation decay (red circles), and our measurements of spurious excitation (blue squares).  The red line is a fit to the Ramsey fringe data using a $1/f^\alpha$ power law, with the grey shaded area representing the 95\% confidence interval on the fitted value of $\alpha=0.57$.  The inset view shows a detail of the Ramsey fringe data and fit line.}
\end{figure}

Given the very weak charge dispersion of the transmon \cite{Koch2007} and the low level of critical current noise in similar sub-micron Josephson junctions \cite{Murch2012}, we expect flux noise to be the dominant source of dephasing in our qubit sample \cite{Note1}.  If we attribute all spurious excitation to up-converted flux noise, we can extrapolate the population data in Fig. \ref{noisefig}(a) back along the flux noise axis until the intercept with the thermal population, yielding an estimate of the intrinsic flux noise at the detuning frequency.  Using this method, we extract a flux noise spectral density  $S_\Phi^{1/2}(705\, \mathrm{MHz})=0.009\pm0.004$ \uphi\rthz.  Using experimentally measured values for $g$, $\Delta_\mathrm{ro}$, $\nu$, $\Gamma_1$, and the slope of $\langle\hat{\sigma}_z\rangle$ versus \nbar, the theoretical dressed dephasing expressions (\ref{ddeq1}) and (\ref{ddeq2}) give a value of $S_\Phi^{1/2}(705\, \mathrm{MHz})=0.011\pm0.002$ \uphi\rthz.  

We performed two other experiments to measure flux noise at other frequencies.  First, we measured Ramsey fringes at a rate of one per second for 90 minutes, then fit each fringe to extract the qubit frequency.  The fluctuations in the qubit frequency over time can be translated into an effective flux noise spectral density for frequencies below 0.5 Hz. Second, we measured Rabi oscillations at different Rabi frequencies $\Omega_R$ and extracted the decay rate $\widetilde{\Gamma}_2$.  We used this decay rate to obtain a spectral density of qubit frequency fluctuations $S_{\delta\omega_q}$ at $\Omega_R$ using the relation $\widetilde{\Gamma}_2=\frac{3}{4}\Gamma_1+\frac{1}{2}\Gamma_\nu$, where $\Gamma_\nu=\pi S_{\delta\omega_q}(\Omega_R)$ \cite{Ithier2005}.  Converting $S_{\delta\omega_q}(\Omega_R)$ to an effective flux noise gives us data for frequencies between 1 and 20 MHz.  

The flux noise values extracted by these methods, as well as those from our spurious excitation data, are shown in Figure \ref{slopefig}.  We fit the Ramsey data to a $1/f^\alpha$ power law \cite{Wellstood1987, Bylander2011} corresponding to the red trend line.  This fit agrees with the extracted values from both Rabi decay and spurious excitation data, representing a power law for flux noise that appears to hold over 11 orders of magnitude in frequency.  The fit coefficients give $\alpha=0.57\pm0.03$ and $S_\Phi^{1/2}(1\,\mathrm{Hz})=1.5\pm0.1\, \mu\Phi_0/\sqrt{\mathrm{Hz}}$, both of which agree with typical values reported in the literature \cite{Wellstood1987, Awschalom1988,Yoshihara2006,Drung2011, Anton2011}.  We note that other recent work produced a similar $S_\Phi^{1/2}(1\,\mathrm{Hz})$ but found $\alpha=0.9-1.0$ \cite{Bylander2011,Bialczak2007}; sample-to-sample variation in $\alpha$ of this magnitude has been noted elsewhere \cite{Wellstood1987,Drung2011, Anton2011}.

The correspondence of the low-frequency fit to the extracted flux noise at $\Delta_\mathrm{ro}(\bar{n})$ suggests that the ``universal'' low-frequency flux noise \cite{Koch2007a} persists to GHz frequencies.  In addition to dephasing the qubit state, this noise also reduces the QND character of circuit QED measurement, in agreement with the dressed dephasing theory.  This sets limits on achievable readout fidelity\textemdash even with quantum-limited post-amplification\textemdash by giving a penalty for increasing \nbar.  Recent efforts to understand and improve low-frequency dephasing noise \cite{Anton2011, Sank2011, Yan2012} may therefore also provide a route to improved qubit readout fidelity.  
\begin{acknowledgments}
We thank O. Naaman, J. Aumentado, and Y. C. Hagar for useful discussions on data processing. D.H.S. acknowledges support from a Hertz Foundation Fellowship endowed by Big George Ventures.  R.V. and I.S. acknowledge support from the Army Research Office QCT program under contract \#W911NF1110029.  We acknowledge support from NSERC (S.B., M.B., A.B.) and FQRNT (M.B.).  A.B. further acknowledges funding from the Alfred P. Sloan Foundation and CIFAR. We thank Calcul Qu\'{e}bec for computation time.
\end{acknowledgments}


\begin{thebibliography}{38}%
\makeatletter
\providecommand \@ifxundefined [1]{%
 \@ifx{#1\undefined}
}%
\providecommand \@ifnum [1]{%
 \ifnum #1\expandafter \@firstoftwo
 \else \expandafter \@secondoftwo
 \fi
}%
\providecommand \@ifx [1]{%
 \ifx #1\expandafter \@firstoftwo
 \else \expandafter \@secondoftwo
 \fi
}%
\providecommand \natexlab [1]{#1}%
\providecommand \enquote  [1]{``#1''}%
\providecommand \bibnamefont  [1]{#1}%
\providecommand \bibfnamefont [1]{#1}%
\providecommand \citenamefont [1]{#1}%
\providecommand \href@noop [0]{\@secondoftwo}%
\providecommand \href [0]{\begingroup \@sanitize@url \@href}%
\providecommand \@href[1]{\@@startlink{#1}\@@href}%
\providecommand \@@href[1]{\endgroup#1\@@endlink}%
\providecommand \@sanitize@url [0]{\catcode `\\12\catcode `\$12\catcode
  `\&12\catcode `\#12\catcode `\^12\catcode `\_12\catcode `\%12\relax}%
\providecommand \@@startlink[1]{}%
\providecommand \@@endlink[0]{}%
\providecommand \url  [0]{\begingroup\@sanitize@url \@url }%
\providecommand \@url [1]{\endgroup\@href {#1}{\urlprefix }}%
\providecommand \urlprefix  [0]{URL }%
\providecommand \Eprint [0]{\href }%
\providecommand \doibase [0]{http://dx.doi.org/}%
\providecommand \selectlanguage [0]{\@gobble}%
\providecommand \bibinfo  [0]{\@secondoftwo}%
\providecommand \bibfield  [0]{\@secondoftwo}%
\providecommand \translation [1]{[#1]}%
\providecommand \BibitemOpen [0]{}%
\providecommand \bibitemStop [0]{}%
\providecommand \bibitemNoStop [0]{.\EOS\space}%
\providecommand \EOS [0]{\spacefactor3000\relax}%
\providecommand \BibitemShut  [1]{\csname bibitem#1\endcsname}%
\let\auto@bib@innerbib\@empty
\bibitem [{\citenamefont {Siddiqi}(2011)}]{Siddiqi2011}%
  \BibitemOpen
  \bibfield  {author} {\bibinfo {author} {\bibfnamefont {I.}~\bibnamefont
  {Siddiqi}},\ }\href {\doibase 10.1088/0953-2048/24/9/091002} {\bibfield
  {journal} {\bibinfo  {journal} {Supercond. Sci. Tech.}\ }\textbf {\bibinfo
  {volume} {24}},\ \bibinfo {pages} {091002} (\bibinfo {year}
  {2011})}\BibitemShut {NoStop}%
\bibitem [{\citenamefont {Clarke}\ and\ \citenamefont
  {Wilhelm}(2008)}]{Clarke2008}%
  \BibitemOpen
  \bibfield  {author} {\bibinfo {author} {\bibfnamefont {J.}~\bibnamefont
  {Clarke}}\ and\ \bibinfo {author} {\bibfnamefont {F.~K.}\ \bibnamefont
  {Wilhelm}},\ }\href {\doibase 10.1038/nature07128} {\bibfield  {journal}
  {\bibinfo  {journal} {Nature}\ }\textbf {\bibinfo {volume} {453}},\ \bibinfo
  {pages} {1031} (\bibinfo {year} {2008})}\BibitemShut {NoStop}%
\bibitem [{\citenamefont {Braginsky}\ and\ \citenamefont
  {Khalili}(1992)}]{Braginsky1992}%
  \BibitemOpen
  \bibfield  {author} {\bibinfo {author} {\bibfnamefont {V.~B.}\ \bibnamefont
  {Braginsky}}\ and\ \bibinfo {author} {\bibfnamefont {F.~Y.}\ \bibnamefont
  {Khalili}},\ }\href@noop {} {\emph {\bibinfo {title} {{Quantum
  Measurement}}}},\ edited by\ \bibinfo {editor} {\bibfnamefont {K.~S.}\
  \bibnamefont {Thorne}}\ (\bibinfo  {publisher} {Cambridge University Press},\
  \bibinfo {address} {New York},\ \bibinfo {year} {1992})\BibitemShut {NoStop}%
\bibitem [{\citenamefont {Blais}\ \emph {et~al.}(2004)\citenamefont {Blais},
  \citenamefont {Huang}, \citenamefont {Wallraff}, \citenamefont {Girvin},\
  and\ \citenamefont {Schoelkopf}}]{Blais2004}%
  \BibitemOpen
  \bibfield  {author} {\bibinfo {author} {\bibfnamefont {A.}~\bibnamefont
  {Blais}}, \bibinfo {author} {\bibfnamefont {R.-S.}\ \bibnamefont {Huang}},
  \bibinfo {author} {\bibfnamefont {A.}~\bibnamefont {Wallraff}}, \bibinfo
  {author} {\bibfnamefont {S.~M.}\ \bibnamefont {Girvin}}, \ and\ \bibinfo
  {author} {\bibfnamefont {R.~J.}\ \bibnamefont {Schoelkopf}},\ }\href
  {\doibase 10.1103/PhysRevA.69.062320} {\bibfield  {journal} {\bibinfo
  {journal} {Phys. Rev. A}\ }\textbf {\bibinfo {volume} {69}},\ \bibinfo
  {pages} {062320} (\bibinfo {year} {2004})}\BibitemShut {NoStop}%
\bibitem [{\citenamefont {Wallraff}\ \emph {et~al.}(2004)\citenamefont
  {Wallraff}, \citenamefont {Schuster}, \citenamefont {Blais}, \citenamefont
  {Frunzio}, \citenamefont {Huang}, \citenamefont {Majer}, \citenamefont
  {Kumar}, \citenamefont {Girvin},\ and\ \citenamefont
  {Schoelkopf}}]{Wallraff2004}%
  \BibitemOpen
  \bibfield  {author} {\bibinfo {author} {\bibfnamefont {A.}~\bibnamefont
  {Wallraff}}, \bibinfo {author} {\bibfnamefont {D.~I.}\ \bibnamefont
  {Schuster}}, \bibinfo {author} {\bibfnamefont {A.}~\bibnamefont {Blais}},
  \bibinfo {author} {\bibfnamefont {L.}~\bibnamefont {Frunzio}}, \bibinfo
  {author} {\bibfnamefont {R.-S.}\ \bibnamefont {Huang}}, \bibinfo {author}
  {\bibfnamefont {J.}~\bibnamefont {Majer}}, \bibinfo {author} {\bibfnamefont
  {S.}~\bibnamefont {Kumar}}, \bibinfo {author} {\bibfnamefont {S.~M.}\
  \bibnamefont {Girvin}}, \ and\ \bibinfo {author} {\bibfnamefont {R.~J.}\
  \bibnamefont {Schoelkopf}},\ }\href {\doibase 10.1038/nature02851} {\bibfield
   {journal} {\bibinfo  {journal} {Nature}\ }\textbf {\bibinfo {volume}
  {431}},\ \bibinfo {pages} {162} (\bibinfo {year} {2004})}\BibitemShut
  {NoStop}%
\bibitem [{\citenamefont {Wallraff}\ \emph {et~al.}(2005)\citenamefont
  {Wallraff}, \citenamefont {Schuster}, \citenamefont {Blais}, \citenamefont
  {Frunzio}, \citenamefont {Majer}, \citenamefont {Devoret}, \citenamefont
  {Girvin},\ and\ \citenamefont {Schoelkopf}}]{Wallraff2005}%
  \BibitemOpen
  \bibfield  {author} {\bibinfo {author} {\bibfnamefont {A.}~\bibnamefont
  {Wallraff}}, \bibinfo {author} {\bibfnamefont {D.~I.}\ \bibnamefont
  {Schuster}}, \bibinfo {author} {\bibfnamefont {A.}~\bibnamefont {Blais}},
  \bibinfo {author} {\bibfnamefont {L.}~\bibnamefont {Frunzio}}, \bibinfo
  {author} {\bibfnamefont {J.}~\bibnamefont {Majer}}, \bibinfo {author}
  {\bibfnamefont {M.~H.}\ \bibnamefont {Devoret}}, \bibinfo {author}
  {\bibfnamefont {S.~M.}\ \bibnamefont {Girvin}}, \ and\ \bibinfo {author}
  {\bibfnamefont {R.~J.}\ \bibnamefont {Schoelkopf}},\ }\href {\doibase
  10.1103/PhysRevLett.95.060501} {\bibfield  {journal} {\bibinfo  {journal}
  {Phys. Rev. Lett.}\ }\textbf {\bibinfo {volume} {95}},\ \bibinfo {pages}
  {060501} (\bibinfo {year} {2005})}\BibitemShut {NoStop}%
\bibitem [{\citenamefont {Hatridge}\ \emph {et~al.}(2011)\citenamefont
  {Hatridge}, \citenamefont {Vijay}, \citenamefont {Slichter}, \citenamefont
  {Clarke},\ and\ \citenamefont {Siddiqi}}]{Hatridge2011}%
  \BibitemOpen
  \bibfield  {author} {\bibinfo {author} {\bibfnamefont {M.}~\bibnamefont
  {Hatridge}}, \bibinfo {author} {\bibfnamefont {R.}~\bibnamefont {Vijay}},
  \bibinfo {author} {\bibfnamefont {D.~H.}\ \bibnamefont {Slichter}}, \bibinfo
  {author} {\bibfnamefont {J.}~\bibnamefont {Clarke}}, \ and\ \bibinfo {author}
  {\bibfnamefont {I.}~\bibnamefont {Siddiqi}},\ }\href {\doibase
  10.1103/PhysRevB.83.134501} {\bibfield  {journal} {\bibinfo  {journal} {Phys.
  Rev. B}\ }\textbf {\bibinfo {volume} {83}},\ \bibinfo {pages} {134501}
  (\bibinfo {year} {2011})}\BibitemShut {NoStop}%
\bibitem [{\citenamefont {Levenson-Falk}\ \emph {et~al.}(2011)\citenamefont
  {Levenson-Falk}, \citenamefont {Vijay},\ and\ \citenamefont
  {Siddiqi}}]{Levenson-Falk2011}%
  \BibitemOpen
  \bibfield  {author} {\bibinfo {author} {\bibfnamefont {E.~M.}\ \bibnamefont
  {Levenson-Falk}}, \bibinfo {author} {\bibfnamefont {R.}~\bibnamefont
  {Vijay}}, \ and\ \bibinfo {author} {\bibfnamefont {I.}~\bibnamefont
  {Siddiqi}},\ }\href {\doibase 10.1063/1.3570693} {\bibfield  {journal}
  {\bibinfo  {journal} {Appl. Phys. Lett.}\ }\textbf {\bibinfo {volume} {98}},\
  \bibinfo {pages} {123115} (\bibinfo {year} {2011})}\BibitemShut {NoStop}%
\bibitem [{\citenamefont {Vijay}\ \emph {et~al.}(2011)\citenamefont {Vijay},
  \citenamefont {Slichter},\ and\ \citenamefont {Siddiqi}}]{Vijay2011}%
  \BibitemOpen
  \bibfield  {author} {\bibinfo {author} {\bibfnamefont {R.}~\bibnamefont
  {Vijay}}, \bibinfo {author} {\bibfnamefont {D.~H.}\ \bibnamefont {Slichter}},
  \ and\ \bibinfo {author} {\bibfnamefont {I.}~\bibnamefont {Siddiqi}},\ }\href
  {\doibase 10.1103/PhysRevLett.106.110502} {\bibfield  {journal} {\bibinfo
  {journal} {Phys. Rev. Lett.}\ }\textbf {\bibinfo {volume} {106}},\ \bibinfo
  {pages} {110502} (\bibinfo {year} {2011})}\BibitemShut {NoStop}%
\bibitem [{\citenamefont {Johnson}\ \emph {et~al.}(2012)\citenamefont
  {Johnson}, \citenamefont {Macklin}, \citenamefont {Slichter}, \citenamefont
  {Vijay}, \citenamefont {Weingarten}, \citenamefont {Clarke},\ and\
  \citenamefont {Siddiqi}}]{Johnson2012}%
  \BibitemOpen
  \bibfield  {author} {\bibinfo {author} {\bibfnamefont {J.~E.}\ \bibnamefont
  {Johnson}}, \bibinfo {author} {\bibfnamefont {C.}~\bibnamefont {Macklin}},
  \bibinfo {author} {\bibfnamefont {D.~H.}\ \bibnamefont {Slichter}}, \bibinfo
  {author} {\bibfnamefont {R.}~\bibnamefont {Vijay}}, \bibinfo {author}
  {\bibfnamefont {E.~B.}\ \bibnamefont {Weingarten}}, \bibinfo {author}
  {\bibfnamefont {J.}~\bibnamefont {Clarke}}, \ and\ \bibinfo {author}
  {\bibfnamefont {I.}~\bibnamefont {Siddiqi}},\ }\href {\doibase
  10.1103/PhysRevLett.109.050506} {\bibfield  {journal} {\bibinfo  {journal}
  {Phys. Rev. Lett.}\ }\textbf {\bibinfo {volume} {109}},\ \bibinfo {pages}
  {050506} (\bibinfo {year} {2012})}\BibitemShut {NoStop}%
\bibitem [{\citenamefont {Rist\`{e}}\ \emph {et~al.}(2012)\citenamefont
  {Rist\`{e}}, \citenamefont {van Leeuwen}, \citenamefont {Ku}, \citenamefont
  {Lehnert},\ and\ \citenamefont {DiCarlo}}]{Riste2012}%
  \BibitemOpen
  \bibfield  {author} {\bibinfo {author} {\bibfnamefont {D.}~\bibnamefont
  {Rist\`{e}}}, \bibinfo {author} {\bibfnamefont {J.~G.}\ \bibnamefont {van
  Leeuwen}}, \bibinfo {author} {\bibfnamefont {H.~S.}\ \bibnamefont {Ku}},
  \bibinfo {author} {\bibfnamefont {K.~W.}\ \bibnamefont {Lehnert}}, \ and\
  \bibinfo {author} {\bibfnamefont {L.}~\bibnamefont {DiCarlo}},\ }\href
  {\doibase 10.1103/PhysRevLett.109.050507} {\bibfield  {journal} {\bibinfo
  {journal} {Phys. Rev. Lett.}\ }\textbf {\bibinfo {volume} {109}},\ \bibinfo
  {pages} {050507} (\bibinfo {year} {2012})}\BibitemShut {NoStop}%
\bibitem [{\citenamefont {Reed}\ \emph {et~al.}(2010)\citenamefont {Reed},
  \citenamefont {DiCarlo}, \citenamefont {Johnson}, \citenamefont {Sun},
  \citenamefont {Schuster}, \citenamefont {Frunzio},\ and\ \citenamefont
  {Schoelkopf}}]{Reed2010a}%
  \BibitemOpen
  \bibfield  {author} {\bibinfo {author} {\bibfnamefont {M.~D.}\ \bibnamefont
  {Reed}}, \bibinfo {author} {\bibfnamefont {L.}~\bibnamefont {DiCarlo}},
  \bibinfo {author} {\bibfnamefont {B.~R.}\ \bibnamefont {Johnson}}, \bibinfo
  {author} {\bibfnamefont {L.}~\bibnamefont {Sun}}, \bibinfo {author}
  {\bibfnamefont {D.~I.}\ \bibnamefont {Schuster}}, \bibinfo {author}
  {\bibfnamefont {L.}~\bibnamefont {Frunzio}}, \ and\ \bibinfo {author}
  {\bibfnamefont {R.~J.}\ \bibnamefont {Schoelkopf}},\ }\href {\doibase
  10.1103/PhysRevLett.105.173601} {\bibfield  {journal} {\bibinfo  {journal}
  {Phys. Rev. Lett.}\ }\textbf {\bibinfo {volume} {105}},\ \bibinfo {pages}
  {173601} (\bibinfo {year} {2010})}\BibitemShut {NoStop}%
\bibitem [{\citenamefont {Boissonneault}\ \emph {et~al.}(2010)\citenamefont
  {Boissonneault}, \citenamefont {Gambetta},\ and\ \citenamefont
  {Blais}}]{Boissonneault2010}%
  \BibitemOpen
  \bibfield  {author} {\bibinfo {author} {\bibfnamefont {M.}~\bibnamefont
  {Boissonneault}}, \bibinfo {author} {\bibfnamefont {J.~M.}\ \bibnamefont
  {Gambetta}}, \ and\ \bibinfo {author} {\bibfnamefont {A.}~\bibnamefont
  {Blais}},\ }\href {\doibase 10.1103/PhysRevLett.105.100504} {\bibfield
  {journal} {\bibinfo  {journal} {Phys. Rev. Lett.}\ }\textbf {\bibinfo
  {volume} {105}},\ \bibinfo {pages} {100504} (\bibinfo {year}
  {2010})}\BibitemShut {NoStop}%
\bibitem [{\citenamefont {Bishop}\ \emph {et~al.}(2010)\citenamefont {Bishop},
  \citenamefont {Ginossar},\ and\ \citenamefont {Girvin}}]{Bishop2010}%
  \BibitemOpen
  \bibfield  {author} {\bibinfo {author} {\bibfnamefont {L.~S.}\ \bibnamefont
  {Bishop}}, \bibinfo {author} {\bibfnamefont {E.}~\bibnamefont {Ginossar}}, \
  and\ \bibinfo {author} {\bibfnamefont {S.~M.}\ \bibnamefont {Girvin}},\
  }\href {\doibase 10.1103/PhysRevLett.105.100505} {\bibfield  {journal}
  {\bibinfo  {journal} {Phys. Rev. Lett.}\ }\textbf {\bibinfo {volume} {105}},\
  \bibinfo {pages} {100505} (\bibinfo {year} {2010})}\BibitemShut {NoStop}%
\bibitem [{\citenamefont {Mallet}\ \emph {et~al.}(2009)\citenamefont {Mallet},
  \citenamefont {Ong}, \citenamefont {Palacios-Laloy}, \citenamefont {Nguyen},
  \citenamefont {Bertet}, \citenamefont {Vion},\ and\ \citenamefont
  {Esteve}}]{Mallet2009}%
  \BibitemOpen
  \bibfield  {author} {\bibinfo {author} {\bibfnamefont {F.}~\bibnamefont
  {Mallet}}, \bibinfo {author} {\bibfnamefont {F.~R.}\ \bibnamefont {Ong}},
  \bibinfo {author} {\bibfnamefont {A.}~\bibnamefont {Palacios-Laloy}},
  \bibinfo {author} {\bibfnamefont {F.}~\bibnamefont {Nguyen}}, \bibinfo
  {author} {\bibfnamefont {P.}~\bibnamefont {Bertet}}, \bibinfo {author}
  {\bibfnamefont {D.}~\bibnamefont {Vion}}, \ and\ \bibinfo {author}
  {\bibfnamefont {D.}~\bibnamefont {Esteve}},\ }\href {\doibase
  10.1038/nphys1400} {\bibfield  {journal} {\bibinfo  {journal} {Nature Phys.}\
  }\textbf {\bibinfo {volume} {5}},\ \bibinfo {pages} {791} (\bibinfo {year}
  {2009})}\BibitemShut {NoStop}%
\bibitem [{\citenamefont {Boissonneault}\ \emph {et~al.}(2009)\citenamefont
  {Boissonneault}, \citenamefont {Gambetta},\ and\ \citenamefont
  {Blais}}]{Boissonneault2009}%
  \BibitemOpen
  \bibfield  {author} {\bibinfo {author} {\bibfnamefont {M.}~\bibnamefont
  {Boissonneault}}, \bibinfo {author} {\bibfnamefont {J.~M.}\ \bibnamefont
  {Gambetta}}, \ and\ \bibinfo {author} {\bibfnamefont {A.}~\bibnamefont
  {Blais}},\ }\href {\doibase 10.1103/PhysRevA.79.013819} {\bibfield  {journal}
  {\bibinfo  {journal} {Phys. Rev. A}\ }\textbf {\bibinfo {volume} {79}},\
  \bibinfo {pages} {013819} (\bibinfo {year} {2009})}\BibitemShut {NoStop}%
\bibitem [{\citenamefont {Wellstood}\ \emph {et~al.}(1987)\citenamefont
  {Wellstood}, \citenamefont {Urbina},\ and\ \citenamefont
  {Clarke}}]{Wellstood1987}%
  \BibitemOpen
  \bibfield  {author} {\bibinfo {author} {\bibfnamefont {F.~C.}\ \bibnamefont
  {Wellstood}}, \bibinfo {author} {\bibfnamefont {C.}~\bibnamefont {Urbina}}, \
  and\ \bibinfo {author} {\bibfnamefont {J.}~\bibnamefont {Clarke}},\ }\href
  {\doibase 10.1063/1.98041} {\bibfield  {journal} {\bibinfo  {journal} {Appl.
  Phys. Lett.}\ }\textbf {\bibinfo {volume} {50}},\ \bibinfo {pages} {772}
  (\bibinfo {year} {1987})}\BibitemShut {NoStop}%
\bibitem [{\citenamefont {Koch}\ \emph
  {et~al.}(2007{\natexlab{a}})\citenamefont {Koch}, \citenamefont
  {DiVincenzo},\ and\ \citenamefont {Clarke}}]{Koch2007a}%
  \BibitemOpen
  \bibfield  {author} {\bibinfo {author} {\bibfnamefont {R.~H.}\ \bibnamefont
  {Koch}}, \bibinfo {author} {\bibfnamefont {D.~P.}\ \bibnamefont
  {DiVincenzo}}, \ and\ \bibinfo {author} {\bibfnamefont {J.}~\bibnamefont
  {Clarke}},\ }\href {\doibase 10.1103/PhysRevLett.98.267003} {\bibfield
  {journal} {\bibinfo  {journal} {Phys. Rev. Lett.}\ }\textbf {\bibinfo
  {volume} {98}},\ \bibinfo {pages} {267003} (\bibinfo {year}
  {2007}{\natexlab{a}})}\BibitemShut {NoStop}%
\bibitem [{\citenamefont {Boissonneault}\ \emph {et~al.}(2012)\citenamefont
  {Boissonneault}, \citenamefont {Doherty}, \citenamefont {Ong}, \citenamefont
  {Bertet}, \citenamefont {Vion}, \citenamefont {Esteve},\ and\ \citenamefont
  {Blais}}]{Boissonneault2012}%
  \BibitemOpen
  \bibfield  {author} {\bibinfo {author} {\bibfnamefont {M.}~\bibnamefont
  {Boissonneault}}, \bibinfo {author} {\bibfnamefont {A.~C.}\ \bibnamefont
  {Doherty}}, \bibinfo {author} {\bibfnamefont {F.~R.}\ \bibnamefont {Ong}},
  \bibinfo {author} {\bibfnamefont {P.}~\bibnamefont {Bertet}}, \bibinfo
  {author} {\bibfnamefont {D.}~\bibnamefont {Vion}}, \bibinfo {author}
  {\bibfnamefont {D.}~\bibnamefont {Esteve}}, \ and\ \bibinfo {author}
  {\bibfnamefont {A.}~\bibnamefont {Blais}},\ }\href
  {http://link.aps.org/doi/10.1103/PhysRevA.85.022305} {\bibfield  {journal}
  {\bibinfo  {journal} {Phys. Rev. A}\ }\textbf {\bibinfo {volume} {85}},\
  \bibinfo {pages} {022305} (\bibinfo {year} {2012})}\BibitemShut {NoStop}%
\bibitem [{\citenamefont {Koch}\ \emph
  {et~al.}(2007{\natexlab{b}})\citenamefont {Koch}, \citenamefont {Yu},
  \citenamefont {Gambetta}, \citenamefont {Houck}, \citenamefont {Schuster},
  \citenamefont {Majer}, \citenamefont {Blais}, \citenamefont {Devoret},
  \citenamefont {Girvin},\ and\ \citenamefont {Schoelkopf}}]{Koch2007}%
  \BibitemOpen
  \bibfield  {author} {\bibinfo {author} {\bibfnamefont {J.}~\bibnamefont
  {Koch}}, \bibinfo {author} {\bibfnamefont {T.~M.}\ \bibnamefont {Yu}},
  \bibinfo {author} {\bibfnamefont {J.}~\bibnamefont {Gambetta}}, \bibinfo
  {author} {\bibfnamefont {A.~A.}\ \bibnamefont {Houck}}, \bibinfo {author}
  {\bibfnamefont {D.~I.}\ \bibnamefont {Schuster}}, \bibinfo {author}
  {\bibfnamefont {J.}~\bibnamefont {Majer}}, \bibinfo {author} {\bibfnamefont
  {A.}~\bibnamefont {Blais}}, \bibinfo {author} {\bibfnamefont {M.~H.}\
  \bibnamefont {Devoret}}, \bibinfo {author} {\bibfnamefont {S.~M.}\
  \bibnamefont {Girvin}}, \ and\ \bibinfo {author} {\bibfnamefont {R.~J.}\
  \bibnamefont {Schoelkopf}},\ }\href {\doibase 10.1103/PhysRevA.76.042319}
  {\bibfield  {journal} {\bibinfo  {journal} {Phys. Rev. A}\ }\textbf {\bibinfo
  {volume} {76}},\ \bibinfo {pages} {042319} (\bibinfo {year}
  {2007}{\natexlab{b}})}\BibitemShut {NoStop}%
\bibitem [{\citenamefont {Khalil}\ \emph {et~al.}(2011)\citenamefont {Khalil},
  \citenamefont {Wellstood},\ and\ \citenamefont {Osborn}}]{Khalil2011}%
  \BibitemOpen
  \bibfield  {author} {\bibinfo {author} {\bibfnamefont {M.~S.}\ \bibnamefont
  {Khalil}}, \bibinfo {author} {\bibfnamefont {F.~C.}\ \bibnamefont
  {Wellstood}}, \ and\ \bibinfo {author} {\bibfnamefont {K.~D.}\ \bibnamefont
  {Osborn}},\ }\href {\doibase 10.1109/TASC.2010.2090330} {\bibfield  {journal}
  {\bibinfo  {journal} {IEEE Trans. Appl. Supercond.}\ }\textbf {\bibinfo
  {volume} {21}},\ \bibinfo {pages} {879} (\bibinfo {year} {2011})}\BibitemShut
  {NoStop}%
\bibitem [{\citenamefont {Geerlings}\ \emph {et~al.}(2012)\citenamefont
  {Geerlings}, \citenamefont {Shankar}, \citenamefont {Edwards}, \citenamefont
  {Frunzio}, \citenamefont {Schoelkopf},\ and\ \citenamefont
  {Devoret}}]{Geerlings2012}%
  \BibitemOpen
  \bibfield  {author} {\bibinfo {author} {\bibfnamefont {K.}~\bibnamefont
  {Geerlings}}, \bibinfo {author} {\bibfnamefont {S.}~\bibnamefont {Shankar}},
  \bibinfo {author} {\bibfnamefont {E.}~\bibnamefont {Edwards}}, \bibinfo
  {author} {\bibfnamefont {L.}~\bibnamefont {Frunzio}}, \bibinfo {author}
  {\bibfnamefont {R.~J.}\ \bibnamefont {Schoelkopf}}, \ and\ \bibinfo {author}
  {\bibfnamefont {M.~H.}\ \bibnamefont {Devoret}},\ }\href {\doibase
  10.1063/1.4710520} {\bibfield  {journal} {\bibinfo  {journal} {Appl. Phys.
  Lett.}\ }\textbf {\bibinfo {volume} {100}},\ \bibinfo {pages} {192601}
  (\bibinfo {year} {2012})}\BibitemShut {NoStop}%
\bibitem [{\citenamefont {Slichter}(2011)}]{Slichter2011}%
  \BibitemOpen
  \bibfield  {author} {\bibinfo {author} {\bibfnamefont {D.~H.}\ \bibnamefont
  {Slichter}},\ }\emph {\bibinfo {title} {{Quantum Jumps and Measurement
  Backaction in a Superconducting Qubit}}},\ \href
  {http://physics.berkeley.edu/research/siddiqi/pubs.html} {Ph.D. thesis},\
  \bibinfo  {school} {Univ. of California, Berkeley} (\bibinfo {year}
  {2011})\BibitemShut {NoStop}%
\bibitem [{\citenamefont {Slichter}\ \emph {et~al.}(2009)\citenamefont
  {Slichter}, \citenamefont {Naaman},\ and\ \citenamefont
  {Siddiqi}}]{Slichter2009}%
  \BibitemOpen
  \bibfield  {author} {\bibinfo {author} {\bibfnamefont {D.~H.}\ \bibnamefont
  {Slichter}}, \bibinfo {author} {\bibfnamefont {O.}~\bibnamefont {Naaman}}, \
  and\ \bibinfo {author} {\bibfnamefont {I.}~\bibnamefont {Siddiqi}},\ }\href
  {\doibase 10.1063/1.3133362} {\bibfield  {journal} {\bibinfo  {journal}
  {Appl. Phys. Lett.}\ }\textbf {\bibinfo {volume} {94}},\ \bibinfo {pages}
  {192508} (\bibinfo {year} {2009})}\BibitemShut {NoStop}%
\bibitem [{\citenamefont {Schuster}\ \emph {et~al.}(2005)\citenamefont
  {Schuster}, \citenamefont {Wallraff}, \citenamefont {Blais}, \citenamefont
  {Frunzio}, \citenamefont {Huang}, \citenamefont {Majer}, \citenamefont
  {Girvin},\ and\ \citenamefont {Schoelkopf}}]{Schuster2005}%
  \BibitemOpen
  \bibfield  {author} {\bibinfo {author} {\bibfnamefont {D.~I.}\ \bibnamefont
  {Schuster}}, \bibinfo {author} {\bibfnamefont {A.}~\bibnamefont {Wallraff}},
  \bibinfo {author} {\bibfnamefont {A.}~\bibnamefont {Blais}}, \bibinfo
  {author} {\bibfnamefont {L.}~\bibnamefont {Frunzio}}, \bibinfo {author}
  {\bibfnamefont {R.-S.}\ \bibnamefont {Huang}}, \bibinfo {author}
  {\bibfnamefont {J.}~\bibnamefont {Majer}}, \bibinfo {author} {\bibfnamefont
  {S.~M.}\ \bibnamefont {Girvin}}, \ and\ \bibinfo {author} {\bibfnamefont
  {R.~J.}\ \bibnamefont {Schoelkopf}},\ }\href {\doibase
  10.1103/PhysRevLett.94.123602} {\bibfield  {journal} {\bibinfo  {journal}
  {Phys. Rev. Lett.}\ }\textbf {\bibinfo {volume} {94}},\ \bibinfo {pages}
  {123602} (\bibinfo {year} {2005})}\BibitemShut {NoStop}%
\bibitem [{\citenamefont {Yuzhelevski}\ \emph {et~al.}(2000)\citenamefont
  {Yuzhelevski}, \citenamefont {Yuzhelevski},\ and\ \citenamefont
  {Jung}}]{Yuzhelevski2000}%
  \BibitemOpen
  \bibfield  {author} {\bibinfo {author} {\bibfnamefont {Y.}~\bibnamefont
  {Yuzhelevski}}, \bibinfo {author} {\bibfnamefont {M.}~\bibnamefont
  {Yuzhelevski}}, \ and\ \bibinfo {author} {\bibfnamefont {G.}~\bibnamefont
  {Jung}},\ }\href {\doibase 10.1063/1.1150519} {\bibfield  {journal} {\bibinfo
   {journal} {Rev. Sci. Instrum.}\ }\textbf {\bibinfo {volume} {71}},\ \bibinfo
  {pages} {1681} (\bibinfo {year} {2000})}\BibitemShut {NoStop}%
\bibitem [{\citenamefont {Murch}\ \emph {et~al.}(2012)\citenamefont {Murch},
  \citenamefont {Weber}, \citenamefont {Levenson-Falk}, \citenamefont {Vijay},\
  and\ \citenamefont {Siddiqi}}]{Murch2012}%
  \BibitemOpen
  \bibfield  {author} {\bibinfo {author} {\bibfnamefont {K.~W.}\ \bibnamefont
  {Murch}}, \bibinfo {author} {\bibfnamefont {S.~J.}\ \bibnamefont {Weber}},
  \bibinfo {author} {\bibfnamefont {E.~M.}\ \bibnamefont {Levenson-Falk}},
  \bibinfo {author} {\bibfnamefont {R.}~\bibnamefont {Vijay}}, \ and\ \bibinfo
  {author} {\bibfnamefont {I.}~\bibnamefont {Siddiqi}},\ }\href {\doibase
  10.1063/1.3700964} {\bibfield  {journal} {\bibinfo  {journal} {Appl. Phys.
  Lett.}\ }\textbf {\bibinfo {volume} {100}},\ \bibinfo {pages} {142601}
  (\bibinfo {year} {2012})}\BibitemShut {NoStop}%
\bibitem [{Note1()}]{Note1}%
  \BibitemOpen
  \bibinfo {note} {The power spectral density of qubit frequency fluctuations
  due to flux noise is estimated to be $\sim 10^4$, $\sim 10^5$, and $\sim
  10^8$ times larger than that from critical current noise \cite {Murch2012},
  capacitance fluctuations \cite {Murch2012, Gao2008a}, and charge noise \cite
  {Ithier2005}, respectively}\BibitemShut {NoStop}%
  \bibitem [{\citenamefont {Gao}\ \emph {et~al.}(2008)\citenamefont {Gao},
  \citenamefont {Daal}, \citenamefont {Martinis}, \citenamefont {Vayonakis},
  \citenamefont {Zmuidzinas}, \citenamefont {Sadoulet}, \citenamefont {Mazin},
  \citenamefont {Day},\ and\ \citenamefont {Leduc}}]{Gao2008a}%
  \BibitemOpen
  \bibfield  {author} {\bibinfo {author} {\bibfnamefont {J.}~\bibnamefont
  {Gao}}, \bibinfo {author} {\bibfnamefont {M.}~\bibnamefont {Daal}}, \bibinfo
  {author} {\bibfnamefont {J.~M.}\ \bibnamefont {Martinis}}, \bibinfo {author}
  {\bibfnamefont {A.}~\bibnamefont {Vayonakis}}, \bibinfo {author}
  {\bibfnamefont {J.}~\bibnamefont {Zmuidzinas}}, \bibinfo {author}
  {\bibfnamefont {B.}~\bibnamefont {Sadoulet}}, \bibinfo {author}
  {\bibfnamefont {B.~A.}\ \bibnamefont {Mazin}}, \bibinfo {author}
  {\bibfnamefont {P.~K.}\ \bibnamefont {Day}}, \ and\ \bibinfo {author}
  {\bibfnamefont {H.~G.}\ \bibnamefont {Leduc}},\ }\href {\doibase
  10.1063/1.2937855} {\bibfield  {journal} {\bibinfo  {journal} {Appl. Phys.
  Lett.}\ }\textbf {\bibinfo {volume} {92}},\ \bibinfo {pages} {212504}
  (\bibinfo {year} {2008})}\BibitemShut {NoStop}%
\bibitem [{\citenamefont {Ithier}\ \emph {et~al.}(2005)\citenamefont {Ithier},
  \citenamefont {Collin}, \citenamefont {Joyez}, \citenamefont {Meeson},
  \citenamefont {Vion}, \citenamefont {Esteve}, \citenamefont {Chiarello},
  \citenamefont {Shnirman}, \citenamefont {Makhlin}, \citenamefont {Schriefl},\
  and\ \citenamefont {Sch\"{o}n}}]{Ithier2005}%
  \BibitemOpen
  \bibfield  {author} {\bibinfo {author} {\bibfnamefont {G.}~\bibnamefont
  {Ithier}}, \bibinfo {author} {\bibfnamefont {E.}~\bibnamefont {Collin}},
  \bibinfo {author} {\bibfnamefont {P.}~\bibnamefont {Joyez}}, \bibinfo
  {author} {\bibfnamefont {P.~J.}\ \bibnamefont {Meeson}}, \bibinfo {author}
  {\bibfnamefont {D.}~\bibnamefont {Vion}}, \bibinfo {author} {\bibfnamefont
  {D.}~\bibnamefont {Esteve}}, \bibinfo {author} {\bibfnamefont
  {F.}~\bibnamefont {Chiarello}}, \bibinfo {author} {\bibfnamefont
  {A.}~\bibnamefont {Shnirman}}, \bibinfo {author} {\bibfnamefont
  {Y.}~\bibnamefont {Makhlin}}, \bibinfo {author} {\bibfnamefont
  {J.}~\bibnamefont {Schriefl}}, \ and\ \bibinfo {author} {\bibfnamefont
  {G.}~\bibnamefont {Sch\"{o}n}},\ }\href {\doibase 10.1103/PhysRevB.72.134519}
  {\bibfield  {journal} {\bibinfo  {journal} {Phys. Rev. B}\ }\textbf {\bibinfo
  {volume} {72}},\ \bibinfo {pages} {134519} (\bibinfo {year}
  {2005})}\BibitemShut {NoStop}%
\bibitem [{\citenamefont {Bylander}\ \emph {et~al.}(2011)\citenamefont
  {Bylander}, \citenamefont {Gustavsson}, \citenamefont {Yan}, \citenamefont
  {Yoshihara}, \citenamefont {Harrabi}, \citenamefont {Fitch}, \citenamefont
  {Cory}, \citenamefont {Nakamura}, \citenamefont {Tsai},\ and\ \citenamefont
  {Oliver}}]{Bylander2011}%
  \BibitemOpen
  \bibfield  {author} {\bibinfo {author} {\bibfnamefont {J.}~\bibnamefont
  {Bylander}}, \bibinfo {author} {\bibfnamefont {S.}~\bibnamefont
  {Gustavsson}}, \bibinfo {author} {\bibfnamefont {F.}~\bibnamefont {Yan}},
  \bibinfo {author} {\bibfnamefont {F.}~\bibnamefont {Yoshihara}}, \bibinfo
  {author} {\bibfnamefont {K.}~\bibnamefont {Harrabi}}, \bibinfo {author}
  {\bibfnamefont {G.}~\bibnamefont {Fitch}}, \bibinfo {author} {\bibfnamefont
  {D.~G.}\ \bibnamefont {Cory}}, \bibinfo {author} {\bibfnamefont
  {Y.}~\bibnamefont {Nakamura}}, \bibinfo {author} {\bibfnamefont {J.-S.}\
  \bibnamefont {Tsai}}, \ and\ \bibinfo {author} {\bibfnamefont {W.~D.}\
  \bibnamefont {Oliver}},\ }\href {\doibase 10.1038/nphys1994} {\bibfield
  {journal} {\bibinfo  {journal} {Nature Phys.}\ }\textbf {\bibinfo {volume}
  {7}},\ \bibinfo {pages} {565} (\bibinfo {year} {2011})}\BibitemShut {NoStop}%
\bibitem [{\citenamefont {Awschalom}\ \emph {et~al.}(1988)\citenamefont
  {Awschalom}, \citenamefont {Rozen}, \citenamefont {Ketchen}, \citenamefont
  {Gallagher}, \citenamefont {Kleinsasser}, \citenamefont {Sandstrom},\ and\
  \citenamefont {Bumble}}]{Awschalom1988}%
  \BibitemOpen
  \bibfield  {author} {\bibinfo {author} {\bibfnamefont {D.~D.}\ \bibnamefont
  {Awschalom}}, \bibinfo {author} {\bibfnamefont {J.~R.}\ \bibnamefont
  {Rozen}}, \bibinfo {author} {\bibfnamefont {M.~B.}\ \bibnamefont {Ketchen}},
  \bibinfo {author} {\bibfnamefont {W.~J.}\ \bibnamefont {Gallagher}}, \bibinfo
  {author} {\bibfnamefont {A.~W.}\ \bibnamefont {Kleinsasser}}, \bibinfo
  {author} {\bibfnamefont {R.~L.}\ \bibnamefont {Sandstrom}}, \ and\ \bibinfo
  {author} {\bibfnamefont {B.}~\bibnamefont {Bumble}},\ }\href {\doibase
  10.1063/1.100291} {\bibfield  {journal} {\bibinfo  {journal} {Appl. Phys.
  Lett.}\ }\textbf {\bibinfo {volume} {53}},\ \bibinfo {pages} {2108} (\bibinfo
  {year} {1988})}\BibitemShut {NoStop}%
\bibitem [{\citenamefont {Yoshihara}\ \emph {et~al.}(2006)\citenamefont
  {Yoshihara}, \citenamefont {Harrabi}, \citenamefont {Niskanen}, \citenamefont
  {Nakamura},\ and\ \citenamefont {Tsai}}]{Yoshihara2006}%
  \BibitemOpen
  \bibfield  {author} {\bibinfo {author} {\bibfnamefont {F.}~\bibnamefont
  {Yoshihara}}, \bibinfo {author} {\bibfnamefont {K.}~\bibnamefont {Harrabi}},
  \bibinfo {author} {\bibfnamefont {A.~O.}\ \bibnamefont {Niskanen}}, \bibinfo
  {author} {\bibfnamefont {Y.}~\bibnamefont {Nakamura}}, \ and\ \bibinfo
  {author} {\bibfnamefont {J.~S.}\ \bibnamefont {Tsai}},\ }\href {\doibase
  10.1103/PhysRevLett.97.167001} {\bibfield  {journal} {\bibinfo  {journal}
  {Phys. Rev. Lett.}\ }\textbf {\bibinfo {volume} {97}},\ \bibinfo {pages}
  {167001} (\bibinfo {year} {2006})}\BibitemShut {NoStop}%
\bibitem [{\citenamefont {Drung}\ \emph {et~al.}(2011)\citenamefont {Drung},
  \citenamefont {Beyer}, \citenamefont {Storm}, \citenamefont {Peters},\ and\
  \citenamefont {Schurig}}]{Drung2011}%
  \BibitemOpen
  \bibfield  {author} {\bibinfo {author} {\bibfnamefont {D.}~\bibnamefont
  {Drung}}, \bibinfo {author} {\bibfnamefont {J.}~\bibnamefont {Beyer}},
  \bibinfo {author} {\bibfnamefont {J.-H.}\ \bibnamefont {Storm}}, \bibinfo
  {author} {\bibfnamefont {M.}~\bibnamefont {Peters}}, \ and\ \bibinfo {author}
  {\bibfnamefont {T.}~\bibnamefont {Schurig}},\ }\href {\doibase
  10.1109/TASC.2010.2084054} {\bibfield  {journal} {\bibinfo  {journal} {IEEE
  Trans. Appl. Supercond.}\ }\textbf {\bibinfo {volume} {21}},\ \bibinfo
  {pages} {340} (\bibinfo {year} {2011})}\BibitemShut {NoStop}%
\bibitem [{\citenamefont {Anton}\ \emph {et~al.}(2012)\citenamefont {Anton},
  \citenamefont {M\"{u}ller}, \citenamefont {Birenbaum}, \citenamefont
  {O'Kelley}, \citenamefont {Fefferman}, \citenamefont {Golubev}, \citenamefont
  {Hilton}, \citenamefont {Cho}, \citenamefont {Irwin}, \citenamefont
  {Wellstood}, \citenamefont {Sch\"{o}n}, \citenamefont {Shnirman},\ and\
  \citenamefont {Clarke}}]{Anton2011}%
  \BibitemOpen
  \bibfield  {author} {\bibinfo {author} {\bibfnamefont {S.~M.}\ \bibnamefont
  {Anton}}, \bibinfo {author} {\bibfnamefont {C.}~\bibnamefont {M\"{u}ller}},
  \bibinfo {author} {\bibfnamefont {J.~S.}\ \bibnamefont {Birenbaum}}, \bibinfo
  {author} {\bibfnamefont {S.~R.}\ \bibnamefont {O'Kelley}}, \bibinfo {author}
  {\bibfnamefont {A.~D.}\ \bibnamefont {Fefferman}}, \bibinfo {author}
  {\bibfnamefont {D.~S.}\ \bibnamefont {Golubev}}, \bibinfo {author}
  {\bibfnamefont {G.~C.}\ \bibnamefont {Hilton}}, \bibinfo {author}
  {\bibfnamefont {H.~M.}\ \bibnamefont {Cho}}, \bibinfo {author} {\bibfnamefont
  {K.~D.}\ \bibnamefont {Irwin}}, \bibinfo {author} {\bibfnamefont {F.~C.}\
  \bibnamefont {Wellstood}}, \bibinfo {author} {\bibfnamefont {G.}~\bibnamefont
  {Sch\"{o}n}}, \bibinfo {author} {\bibfnamefont {A.}~\bibnamefont {Shnirman}},
  \ and\ \bibinfo {author} {\bibfnamefont {J.}~\bibnamefont {Clarke}},\ }\href
  {\doibase 10.1103/PhysRevB.85.224505} {\bibfield  {journal} {\bibinfo
  {journal} {Phys. Rev. B}\ }\textbf {\bibinfo {volume} {85}},\ \bibinfo
  {pages} {224505} (\bibinfo {year} {2012})}\BibitemShut {NoStop}%
\bibitem [{\citenamefont {Bialczak}\ \emph {et~al.}(2007)\citenamefont
  {Bialczak}, \citenamefont {McDermott}, \citenamefont {Ansmann}, \citenamefont
  {Hofheinz}, \citenamefont {Katz}, \citenamefont {Lucero}, \citenamefont
  {Neeley}, \citenamefont {O'Connell}, \citenamefont {Wang}, \citenamefont
  {Cleland},\ and\ \citenamefont {Martinis}}]{Bialczak2007}%
  \BibitemOpen
  \bibfield  {author} {\bibinfo {author} {\bibfnamefont {R.~C.}\ \bibnamefont
  {Bialczak}}, \bibinfo {author} {\bibfnamefont {R.}~\bibnamefont {McDermott}},
  \bibinfo {author} {\bibfnamefont {M.}~\bibnamefont {Ansmann}}, \bibinfo
  {author} {\bibfnamefont {M.}~\bibnamefont {Hofheinz}}, \bibinfo {author}
  {\bibfnamefont {N.}~\bibnamefont {Katz}}, \bibinfo {author} {\bibfnamefont
  {E.}~\bibnamefont {Lucero}}, \bibinfo {author} {\bibfnamefont
  {M.}~\bibnamefont {Neeley}}, \bibinfo {author} {\bibfnamefont {A.~D.}\
  \bibnamefont {O'Connell}}, \bibinfo {author} {\bibfnamefont {H.}~\bibnamefont
  {Wang}}, \bibinfo {author} {\bibfnamefont {A.~N.}\ \bibnamefont {Cleland}}, \
  and\ \bibinfo {author} {\bibfnamefont {J.~M.}\ \bibnamefont {Martinis}},\
  }\href {\doibase 10.1103/PhysRevLett.99.187006} {\bibfield  {journal}
  {\bibinfo  {journal} {Phys. Rev. Lett.}\ }\textbf {\bibinfo {volume} {99}},\
  \bibinfo {pages} {187006} (\bibinfo {year} {2007})}\BibitemShut {NoStop}%
\bibitem [{\citenamefont {Sank}\ \emph {et~al.}(2012)\citenamefont {Sank},
  \citenamefont {Barends}, \citenamefont {Bialczak}, \citenamefont {Chen},
  \citenamefont {Kelly}, \citenamefont {Lenander}, \citenamefont {Lucero},
  \citenamefont {Mariantoni}, \citenamefont {Megrant}, \citenamefont {Neeley},
  \citenamefont {O'Malley}, \citenamefont {Vainsencher}, \citenamefont {Wang},
  \citenamefont {Wenner}, \citenamefont {White}, \citenamefont {Yamamoto},
  \citenamefont {Yin}, \citenamefont {Cleland},\ and\ \citenamefont
  {Martinis}}]{Sank2011}%
  \BibitemOpen
  \bibfield  {author} {\bibinfo {author} {\bibfnamefont {D.}~\bibnamefont
  {Sank}}, \bibinfo {author} {\bibfnamefont {R.}~\bibnamefont {Barends}},
  \bibinfo {author} {\bibfnamefont {R.~C.}\ \bibnamefont {Bialczak}}, \bibinfo
  {author} {\bibfnamefont {Y.}~\bibnamefont {Chen}}, \bibinfo {author}
  {\bibfnamefont {J.}~\bibnamefont {Kelly}}, \bibinfo {author} {\bibfnamefont
  {M.}~\bibnamefont {Lenander}}, \bibinfo {author} {\bibfnamefont
  {E.}~\bibnamefont {Lucero}}, \bibinfo {author} {\bibfnamefont
  {M.}~\bibnamefont {Mariantoni}}, \bibinfo {author} {\bibfnamefont
  {A.}~\bibnamefont {Megrant}}, \bibinfo {author} {\bibfnamefont
  {M.}~\bibnamefont {Neeley}}, \bibinfo {author} {\bibfnamefont {P.~J.~J.}\
  \bibnamefont {O'Malley}}, \bibinfo {author} {\bibfnamefont {A.}~\bibnamefont
  {Vainsencher}}, \bibinfo {author} {\bibfnamefont {H.}~\bibnamefont {Wang}},
  \bibinfo {author} {\bibfnamefont {J.}~\bibnamefont {Wenner}}, \bibinfo
  {author} {\bibfnamefont {T.~C.}\ \bibnamefont {White}}, \bibinfo {author}
  {\bibfnamefont {T.}~\bibnamefont {Yamamoto}}, \bibinfo {author}
  {\bibfnamefont {Y.}~\bibnamefont {Yin}}, \bibinfo {author} {\bibfnamefont
  {A.~N.}\ \bibnamefont {Cleland}}, \ and\ \bibinfo {author} {\bibfnamefont
  {J.~M.}\ \bibnamefont {Martinis}},\ }\href {\doibase
  10.1103/PhysRevLett.109.067001} {\bibfield  {journal} {\bibinfo  {journal}
  {Phys. Rev. Lett.}\ }\textbf {\bibinfo {volume} {109}},\ \bibinfo {pages}
  {067001} (\bibinfo {year} {2012})}\BibitemShut {NoStop}%
\bibitem [{\citenamefont {Yan}\ \emph {et~al.}(2012)\citenamefont {Yan},
  \citenamefont {Bylander}, \citenamefont {Gustavsson}, \citenamefont
  {Yoshihara}, \citenamefont {Harrabi}, \citenamefont {Cory}, \citenamefont
  {Orlando}, \citenamefont {Nakamura}, \citenamefont {Tsai},\ and\
  \citenamefont {Oliver}}]{Yan2012}%
  \BibitemOpen
  \bibfield  {author} {\bibinfo {author} {\bibfnamefont {F.}~\bibnamefont
  {Yan}}, \bibinfo {author} {\bibfnamefont {J.}~\bibnamefont {Bylander}},
  \bibinfo {author} {\bibfnamefont {S.}~\bibnamefont {Gustavsson}}, \bibinfo
  {author} {\bibfnamefont {F.}~\bibnamefont {Yoshihara}}, \bibinfo {author}
  {\bibfnamefont {K.}~\bibnamefont {Harrabi}}, \bibinfo {author} {\bibfnamefont
  {D.~G.}\ \bibnamefont {Cory}}, \bibinfo {author} {\bibfnamefont {T.~P.}\
  \bibnamefont {Orlando}}, \bibinfo {author} {\bibfnamefont {Y.}~\bibnamefont
  {Nakamura}}, \bibinfo {author} {\bibfnamefont {J.-S.}\ \bibnamefont {Tsai}},
  \ and\ \bibinfo {author} {\bibfnamefont {W.~D.}\ \bibnamefont {Oliver}},\
  }\href {\doibase 10.1103/PhysRevB.85.174521} {\bibfield  {journal} {\bibinfo
  {journal} {Phys. Rev. B}\ }\textbf {\bibinfo {volume} {85}},\ \bibinfo
  {pages} {174521} (\bibinfo {year} {2012})}\BibitemShut {NoStop}%

\end{thebibliography}

%

\end{document}